\documentstyle[prb,aps,twocolumn,epsfig]{revtex}
\begin{document}
\title{From spinons to magnons in explicit and spontaneously
dimerized antiferromagnetic chains}
\author{Ariel Dobry}
\address{Departamento de F\'{\i}sica, Universidad Nacional de Rosario, \\
and Instituto de F\'{\i}sica Rosario, Avenida Pellegrini 250,\\
2000 Rosario, Argentina}
\author{David Ibaceta}
\address{Instituto de Astronom\'{\i}a y F\'{\i}sica del Espacio,\\
Casilla de Correos 67, Sucursal 28,\\
1428 Buenos Aires, Argentina.} 
\maketitle
\begin{abstract}
We reconsider the excitation spectra of a dimerized and
 frustrated antiferromagnetic Heisenberg chain. This model is 
taken as the simpler example of compiting spontaneous and
explicit dimerization relevant for 
Spin-Peierls compounds. 
The bosonized theory is a two frequency Sine-Gordon field theory.
We analize the excitation spectrum by semiclassical methods.
The elementary triplet excitation corresponds to an extended magnon whose
radius diverge for vanishing dimerization.
The internal oscilations of the magnon give rise to a series
of excited state until another magnon is emited and a two magnon
continuum is reached. We discuss, for weak
dimerization, in which way the magnon forms as a result
 of a spinon-spinon interaction potential.  
\end{abstract}

One dimensional antiferromagnetic systems have attracted a great deal
of interest over the last decades. They 
 are typical examples where the low 
dimensionality enhances quantum fluctuations producing effects
 completely different than the ones expected by classical
 theories of magnetism.
The interest in one-dimensional magnetic system have recently been 
strongly renewed  in view of the discovery of  a new class 
, in general inorganic compounds, with  definite
one-dimensional character . Detailed spectral characterizations 
are now avaible and the theory is pushed to make precise predictions
on  how different microscopic interactions influence the spectral
response of this system.
We discuss in this work the case of a gapped  
antiferomagnetic chain and analyze how an explicit  dimerization and
a frustrated interaction
compete between them 
and how the magnetic spectra evolve with the microscopic
parameters.
A similar subject has 
recently been analized by different numerical technics
\cite{Sorensen},\cite{Bouze}.
 An analytical approach
for the case of weak dimerization limit has been also proposed
\cite{Affleck}. We will discuss the similarities and differences
with our results.
We take the recently discovered inorganic material CuGeO$_3$
\cite{Hase} as a reference even though our result could be more general.
Our started model Hamiltonian reads: 
\begin{eqnarray}
\frac{H}J=\sum_i\left\{(1+\delta (-1)^i) {\bf S}_i \cdot 
{\bf S}_{i+1}+\alpha {\bf S}_i \cdot {\bf S}_{i+2}\right\}  
\label{H}
\end{eqnarray}
where $J$ is the nearest-neighbor (NN) exchange coupling,  $\alpha$
the frustration parameter and $\delta$ measures the amount of the explicit
dimerization.
There are two  different motivations to associate this model-Hamiltonian
with the magnetic spectra of a Spin-Peierls compound.
First, the temperature dependence of the  magnetic 
susceptibility of CuGeO$_3$ could only be accounted for  if an important 
next nearest-neighbor (NNN) is included in the 1D
Heisenberg model used to describe this material.\cite{RieraD,Castilla}
It is therefore natural to include an explicit dimerization on the
exchange interaction
to describe the magnetic excitations in the low temperature
Peierls phase
and this has been in fact the approach used in various theoretical
studies \cite{RieraKo},\cite{UhligS}.   
Second, if  the adiabatic hypothesis
is left out, the system turn to be a coupled spin-phonon one.
Even though the magnetic interactions have mainly
 one-dimensional character they
interact with the three dimensional phonons. The excitations of the 
low temperature phase are strongly affected by the interchain elastic
coupling. In fact, topological solitons (kink) which are genuine excitations
of the isolated chain cannot longer exist as free excitations because 
they create a zone with opposite dimerization phase relative to the surrounding
 chains. A kink and an antikink are strongly interacting excitations. 
Their interactions were accounted in previous work by a linear confined
potential \cite{Affleck,Augier} or as producing a kind of domain configuration
where the kink and antikink oscillate around an equilibrium distance
\cite{DoIba}. As it has been recently remarked
the  model Hamiltonian (\ref{H}) is the simplest example 
where this processes could be studied.    
In this context the first term of Eq. (\ref{H}) account for the explicit
dimerization imposed by neighboring chains and the second term for
the tendency to spontaneous dimerization as we will discuss in the following.  

For $\alpha=0$ and strong dimerization ($\delta$)
the ground state is
a product of singlet over the strong bonds. The first excitations
 correspond to promote
a singlet to a triplet and then delocalize it to build states of definite
momentum. They form  a band of spin-1 magnon excitations.
 In the opposite limit
($\delta=0$) translational symmetry is not explicitly broken. However for 
$\alpha$ greater than a critical value ($\alpha_c \sim 0.23$) the system
 spontaneously
dimerize and a gap in the spectra is open. At   
$\alpha=1/2$ (Majumdar-Ghosh point) the exact ground state
 is the double degenerate product of singlets
over the nn bonds. 
The excitations have been variationally evaluated by Shastry and Sutherland
\cite{SS} (SS). 
They are massive S=1/2 spinons which correspond to an uncoupled spin separating 
two regions of singlet dimers.
 The two previously discussed 
cases represent the extremal situation of correlation
 length equal to the lattice
constant. As we discuss below the structure of the lower energy triplet
excitations remains 
similar for other values of the correlation length  in the limits previoulsy
discussed.

The main question we address in this work is how the spectra evolve from 
one limit to another i.e. how the massive spinons bound in a magnon.
To address this question we analyze this system in the
limit of  small $\delta$ and
 $\alpha$ slightly larger than the critical value. In this
region  the low energy spectrum
could be studied by bosonization. The bosonized Hamiltonian is
\cite{Affleck2}:

\begin{eqnarray}
H_{bos}=\int dx\{ \frac{u \beta^2}2 \Pi^2+\frac{u}{2 \beta^2}
(\partial_x\phi )^2+g_1 \sin \phi\nonumber\\ 
+g_2 \cos (2\phi )\}
\label{2sin}
\end{eqnarray}
$g_1 \sim \delta$ and $g_2 \sim \alpha-\alpha_c$.    
The proportionality constants are of the order of the unity 
depending on the short range cutoff in the bosonization
procedure.$\beta=\sqrt{2\pi}$ for the isotropic model and $u$
is the spin wave velocity.
The spectrum of this double frequency sine-Gordon (DFSG) field theory
 is not known in general.
We start by analyzing  both limits of 
$g_1=0$ and  $g_2=0$ where the theory reduces to a single Sine-Gordon 
and a lot of information is disponible.
In particular the semiclassical method of Dashen, Hasslasher and Neveu
\cite{DHN} gives the exact mass spectrum of the particles for this model.

For $g_2=0$ (we are indeed analyzing the whole zone $g_2 < 0$
because this marginal interaction  
renormalize to zero in this case) 
  the excitation spectra consists of a
 kink, an antikink and two breathers.
In the original spin language,
the kink excitation carries $S_z=1$, the antikink $S_z=-1$
 and the breathers
$S_z=0$. The lower energy breather is degenerate with the kink 
 and antikink
and these three excitations give rise to a triplet branch, the correlate
of the previously discussed magnon band.
 For $g_1=0$ the only one particle excitations of the resulting SG theory 
are a kink and 
an antikink, no breather is found. This kink carries spin one-half
and is the analog of the massive spinons of the SS variational wave function.
However note  that in this parameter regime 
where the continuum approximation is reliable  
the characteristic
width of these excitations is large compared with the lattice space.      

We are now in  position to discuss the excitations in the intermediate 
zone. We start by implementing a semiclassical calculation of the spectra
of the theory (\ref{2sin}).
 Even though the method is not expected to give exact
results for this non integrable model, it gives a valuable non perturbative 
information which will allow us to interpolate between the desired limits.
The time-independent equation of motion corresponding to (\ref{2sin}) is:
\begin{eqnarray}
\frac{u}{2 \beta^2}(\partial _x ^2 \phi_0)+g_1 \cos\phi_0 - 2 g_2
\sin2\phi_0 = 0
\label{cleq}
\end{eqnarray}
The lowest energy (homogeneous) configuration is 
 $\phi_0=-\frac{\pi} 2\; (\rm{mod}\; 2\pi)$
 (it is not degenerate with  $\phi_0=\frac{\pi}2$ as it would if 
$g_1=0$).
Solitons are  x-dependent solutions of the equation of motion 
 with finite energy with  respect to this homogeneous solution.
Owing to the fact that
 the system is Lorentz invariant, moving solitons are obtained
from the static ones by a Lorentz transformation. 
To look for these static configurations we write a first integral
of Eq. (\ref{cleq}):
\begin{eqnarray}
\frac{(1+B)\xi^2}{2}(\partial_x  \phi_0)^2 - B \sin\phi_0 - \frac{1}4 
\cos2\phi_0 = cte
\label{1int}
\end{eqnarray}
where we have introducted the quantity $B=g_1/(4 g_2)$
as a characteristic parameter to analize the evolution of the spectrum
from one limit to another.
$\xi=u/\Delta$ and $\Delta=\beta \sqrt{u(g_1+4 g_2)}$
 is the mass gap to create a particle (let it call a meson) 
above the homogeneous
vacuum. $\xi$ is the correlation length of the theory and measure
the width of the kinks.   
The boundary conditions for finite energy are 
$\lim_{x \rightarrow \pm \infty} \phi_0 = -\frac{\pi}2\; (\rm{mod}\; 2\pi)$ and    
$\lim_{x \rightarrow \pm \infty} \partial_x\phi_0 = 0$.
They fix the $cte=B+\frac{1}4$. Equation (\ref{1int}) could now be integrated.
 The solution center at the origin is:
\begin{eqnarray}
t(x)\equiv
\sin (\phi_0)= 1\;\;-\;\;2\;\; {\cosh}^2 [x_0/\xi] \;\; {\rm %
sech} [( x-x_0)/\xi]  \nonumber \\
{\rm sech} [( x+x_0 )/ \xi] 
  \label{t}
\end{eqnarray}
with 
\begin{eqnarray}
x_0=\frac12 \log\left[\frac{2+B+\sqrt{1+B}}B\right]  
\label{x0}
\end{eqnarray}   
The quantity $x_0$ define a length scale 
 which can be identified as the 'radius'
of this excitations.  
In the limit of  
$B << 1$ we have $x_0 >> 1$ and $t(x)$ could be approximated by
$\tanh(x-x_0) \tanh(x+x_0)$, i.e. a kink-antikink pair  
of the system without explicit dimerization. In the opposite limit $x_0=0$ and 
(\ref{t}) becomes a kink solution of the static sine-Gordon equation
 corresponding to the dimerized chain. So that, $t(x)$ inter-poles
 between this limiting cases. 

$\phi_0(x)=\rm{arcsin}[ t(x)]$ winds a complete round clockwise
 or counter-clockwise. Therefore this excitation carries $S_z=\pm1$.
The $S_z=0$ component of this triplet excitation 
could be taken into account in our semiclassical approach by including 
the periodic time-dependent solution of the field equation which cannot be
obtained by a boosting of the static solution $t(x)$. 
After quantization, this solution  will  give additional singlet states
which are  not taken into account in the quantum states generated by
the quantization of $t(x)$.
For the single Sine Gordon equation these are the breather-like solutions.
Furthermore, DHN have shown that
the first breather is nothing but a renormalized meson.
In this context the n-excited breather state is considered as bound state
of n mesons.  
For our DFSG equation there are not analytic expressions for these 
breather solutions. We will consider in the following
that a renormalized meson correspond  
to the $S_z=0$ component of our triplet excitation. That is to say, for
the SU(2) invariant model we should expect that after an appropriated
resumation of  
the perturbative series the meson mass $\Delta$ will be equal to the
soliton mass to be introduced below.

\begin{figure}[htb]

\epsfig{file=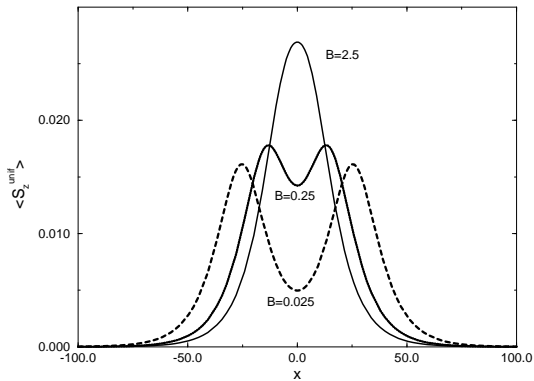,width=6cm,angle=-90}

\vskip .5 truecm

\caption{ Uniform part of the local magnetization carries by
the configuration defined by t(x) for three values of the parameter B}

\label{fig1}
\end{figure}

To discuss the physical content of solution \ref{t} we 
show in Fig. (\ref{fig1}) 
the unstagered
part of the local magnetization $<S^z_i>$  ($\frac{1}{2 \pi}
\partial_x \phi$ in the bosonic representation) for different values 
of the ratio of the parameters B corresponding to $g_1$ smaller, equal or
greater than $g_2$.   
 The correlation length($\xi$) has been fixed to be 
ten times the lattice constant in this Figure.
The figure have been obtained by freezing t(x). This corresponds to
the physical situation where two nonmagnetic impurities cut the chain
and fix the dimerization phase at the border.
This situation  
 have been recently studied by different numerical technics\cite{Sorensen}.
Note the similarities of the behavior showed in Fig. 1 and the
one of Fig. 3a of ref. \onlinecite{Sorensen}.
 For the clean translational invariant
system the local magnetization is x-independent as the result of the 
combination of this pattern
center at the different site of the chain.  
Only for very  low  
B there are two well separated peaks corresponding to two
spin-1/2 excitations
at a distance of the order $x_0/\xi$. For increasing B this
two peaks are washed up and a single spin one excitation emerges at the
chain center. All  the intermediate spins in this configuration
are excited from their value in the ground state.   
The excitation we are dealing up is therefore an extended spin-1 magnon
which dissociate in two independent spinons for very weak dimerization
and collapse in the magnon of the dimerized chain in the opposite
strong dimerization limit.   
Note that at this leading order of our semiclassical expansion this is 
a rigid configuration which translates as a whole.

\begin{figure}[htb]
\epsfig{file=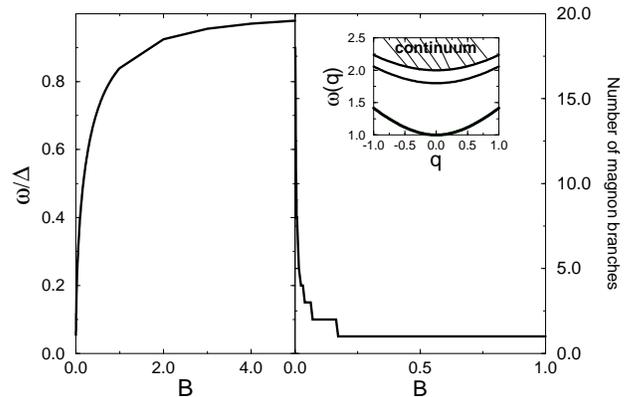,width=6cm,angle=-90}

\vskip .5 truecm

\caption{(a) The evolution of the non-zero eigenfrequency of the fluctuation
operator as obtained by numerical integration of (\ref{scheq}).
(b) The number of  magnon branches predicted by the
semiclassical calculation. The insertion shows a typical low 
energy spectrum.
}

\label{fig2}
\end{figure}

We now include gaussian fluctuation around  this classical
solution. 
The eigenvalues problem of the fluctuation operator
corresponds to a Schroedinger-like
 equation which reads as:

\begin{eqnarray}
[-\partial_x ^2 + V(x)]\psi=E \psi\nonumber \\
V(x)=-\frac{1}{1+B}[B t + (1-2 t(x)^2)] 
\label{scheq}
\end{eqnarray}
where we have made the substitution $x\rightarrow\frac{x}{\xi}$.
The eigenfrequencies are  given by
$\omega=\Delta\sqrt{E}$.
 Equation (\ref{scheq}) has
a zero energy solution (with eigenfunction $\psi=\partial_x\phi_0$)
, this is connected with the
 translational mode arising in the broken of translational
invariance of $t(x)$.
We recall that this is the only bound state of the fluctuation operator
for the case of the single sine-Gordon solitons. 
For the present case we find in addition another bound state which split from   
this zero mode for finite $B$. For increasing $B$ the
energy of this mode increases going to the continuum spectra of 
$V(x)$ for $B \rightarrow \infty$.
In Fig. \ref{fig2} we show the evolution of this eigenvalue
 with B. This figure has been obtained by numerical integration of 
Eq. (\ref{scheq}).

 The excitation spectra of  
the theory in the sector of $Sz=\pm 1$ is spanned by the following
states:
\begin{itemize}

\item
The state of the quantum particle (of mass M)
builded around $t(x)$.
We identify this as belonging to a one-magnon branch
with dispersion $E(p)=u\sqrt{p^2+M^2}$

\item
The excited states of the magnon of mass $M^*=M+n\omega$.
There are as much as adittional triplet branches as $n \omega$ reaches 
the value $\Delta$ where the continuum of the Schroedinger equation starts.
This continuum corresponds to:

\item Labeling by q the continuum of level of Eq.
(\ref{scheq}), they are $\omega(q)=u \sqrt{q^2+\frac{1}{\xi^2}}$  
This is just the kinetic energy of a meson with momentum
q. Therefore this state corresponds to the scattering of a meson
in presence of the soliton. Moreover,
 when one of the continuum modes is excited once, we
get a two particle meson-soliton state. As we have associated a meson
with the elementary $S_z=0$ excitation of the theory
these excitations correspond to the $S_z=1$ 
component of the two magnon continuum of our original spin chain.    

\end{itemize}

The semiclassical calculation predicts the apparition of additional
equi-spaciated magnon branches. The number of these additional branches
diverges as B goes to zero; this is nothing but the two spinons continuum
of the undimerized chain. For decreasing dimerization or increasing
frustration the number of the excited magnon states decrease until
value of B. Beyond this critical value only
 two magnon branches are found. In Fig. \ref{fig2} (b)
we show the behavior previously discussed. The insertion gives our
prediction for the low energy spectra for B greater than the critical value.
Previous exact diagonalization studies of the spectra of Hamiltonian
\ref{H} have  shown the apparition of an additional triplet branch
\cite{RieraKo,Bouze} confirming this picture. The increase of the number of 
the triplet branches with $\delta$ has also recently seen in numerical
studies of this system\cite{Sorensen}. 

Now, we address the question of the contact of this semiclassical approach
and previous work assuming a linear confinement potential between the spinons.
  
For $B<<1$ (weak dimerization limit)
  it is customary to think the problem as the one of two
interacting kinks as has been recently proposed\cite{Affleck,Sorensen}.
 If they are well separated their interaction
is given by an attractive linear potential, the one arising on the term
proportional to $g_1$ in Eq. (\ref{2sin}). At low energy the spectra
of the system corresponds to the one of two particle interacting via
this linear potential. A ladder of bound states is obtained by solution
of an effective Schroedinger equation. At high enough energy these 
bound states are truncated because it becomes favorable to create a 
new kink-antikink pair than to excite one of them. This is essentially
the scenario discussed in ref. (\onlinecite{Affleck})    

\begin{figure}[htb]

\epsfig{file=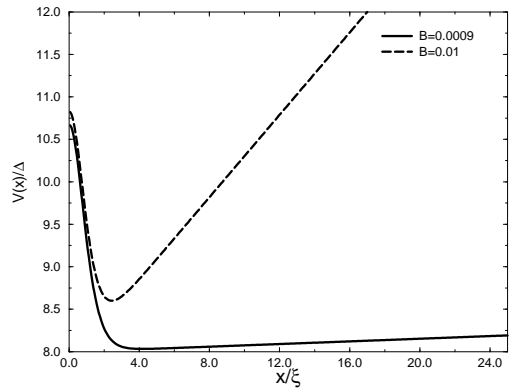,width=6cm,angle=-90}

\vskip .5 truecm

\caption{ 
kink-antikink interaction potential as defined in the text}

\label{fig3}
\end{figure}

Let us define an effective kink-antikink potential by the following
 procedure:
 take Eq.(\ref{t})
 as a suitable 
 kink-antikink form with $x_0$ a variable distance,
replace this form in Eq. (\ref{2sin}) so the $x_0$-dependence of the 
total energy gives our kink-antikink 
interaction energy. The result is show in Fig. (\ref{fig3}) for two
small values of B where the kinks are expected to be well defined excitations.

Two competing interactions could be clearly identify. An attractive linear
interaction and a repulsive one acting at small distance.
The repulsive interaction is originated on the finite width of the kinks.
 The value of
$x_0$ given in eq. (\ref{x0}) represent the minimuun of the total 
potential. For very small B the slope of the linear potential is small
and this minimum is not well defined. The kinks and antikinks will
be always far apart and they never  seen the barrier. Therefore the 
linear approximation for the kink-antikink potential will give
essentially the same result as a potential like the one showed in the figure.
The semiclassical approach gives only poor quantitative results in this
 zone because
it describes the problem as the one of an harmonic oscillators with very small
frequency. However the general qualitative behavior of the spectra
is well reproduced, i.e. the divergence of the number of excited states in 
the limit of no dimerization.  

For increasing B, a well defined minimum appears in the potential energy.
 Now, at low energy the kink-antikink oscillates around this minimum as 
bounded by a spring. 
Our system behaves as a vibrating molecule made by a kink and an antikink.
The semiclassical approximation previously discussed 
precisely describes the quantum state of this harmonic oscillator.

By further increasing B the kinks loose their identity, t(x) becomes  
almost a kink solution of the SG theory with $g_2=0$. The semiclassical
result represents an internal oscilation mode in which the kink waveform
undergoes a harmonically varying shape change localized about the
 kink center. 

Note that even for very small B the kinks are not well separated in
the order of their width, thererofe we should associate the excitation
in a broad range of the parameter as a unique magnon dispersed over a 
great number of spins.
Finally we analyze the experimental situation in CuGeO$_3$.
Neutron scattering measurements of the magnetic spectra
 \cite{Ain} shows only one dispersive excitation before the continuum.
The excited states previously discussed has not been found in these experiments.
Two possible reasons could be given. As it has recently show
 by numerical diagonalization \cite{Sorensen}
the spectral weight of these states are very small.
Moreover, in the original spin-phonon model the inclusion
of the interchain elastic coupling\cite{DoIba} implies that spins interact
with transversal acoustic phonons. Therefore this excited state could
decay in acoustic phonons so broadening the magnon peak.
More experimental and theoretical work will be needed to elucidate 
this point.

We thank I. Affleck for useful discussions and J. Riera 
 for  critical reading of the manuscript.
 A. D. acknowledges   
Fundaci\'on Antorchas for financial support.

\end{document}